\documentclass[english,aps,prd,twocolumn,superscriptaddress,preprintnumbers,nofootinbib]{revtex4}
\usepackage[T1]{fontenc}
\usepackage[latin9]{inputenc}
\usepackage{verbatim}

\usepackage{empheq}
\usepackage{amssymb}
\usepackage{amsmath}
\usepackage{amsfonts}

\usepackage{hyperref}

\newcommand{\be}{\begin{equation}}
\newcommand{\ee}{\end{equation}}

\usepackage{xcolor}
\definecolor{darkblue}{rgb}{0,0,0.7}

\usepackage{dcolumn}
\usepackage{graphicx}


\usepackage{babel}
\usepackage{color}
\begin{document}
\global\long\def\order#1{\mathcal{O}\left(#1\right)}
\global\long\def\d{\mathrm{d}}
\global\long\def\P{P}
\global\long\def\amp{{\mathcal M}}
\preprint{TTP21-006, P3H-21-013, OUTP-21-05P, CERN-TH-2021-027}

\def \SS {S{\hspace{-5pt}}S}
\def\KIT{Institute for Theoretical Particle Physics, KIT, Karlsruhe, Germany}
\def\CERN{Theoretical Physics Department, CERN, 1211 Geneva 23, Switzerland}
\def\OX{Rudolf Peierls Centre for Theoretical Physics, Clarendon Laboratory, Parks Road, Oxford OX1 3PU, UK}
\def\Wadham{Wadham College, University of Oxford, Parks Rd, Oxford OX1 3PN, UK }

\title{Estimating the impact of mixed QCD-electroweak corrections on
  the $W$-mass determination at the LHC}

\author{Arnd  Behring}            
\email[Electronic address: ]{arnd.behring@kit.edu}
\affiliation{\KIT}

\author{Federico Buccioni}            
\email[Electronic address: ]{federico.buccioni@physics.ox.ac.uk}
\affiliation{\OX}

\author{Fabrizio Caola}            
\email[Electronic address: ]{fabrizio.caola@physics.ox.ac.uk}
\affiliation{\OX}
\affiliation{\Wadham}

\author{Maximilian Delto}            
\email[Electronic address: ]{maximilian.delto@kit.edu}
\affiliation{\KIT}

\author{Matthieu Jaquier}            
\email[Electronic address: ]{matthieujaq@gmail.com}
\affiliation{\KIT}

\author{Kirill Melnikov}            
\email[Electronic address: ]{kirill.melnikov@kit.edu}
\affiliation{\KIT}

\author{Raoul R\"ontsch }            
\email[Electronic address: ]{raoul.rontsch@cern.ch}
\affiliation{\CERN}

\begin{abstract}
We study the impact of the recently computed mixed QCD-electroweak
corrections to the production of $W$ and $Z$ bosons at the LHC on the
value of the $W$ mass extracted from the transverse momentum
distribution of charged leptons from $W$ decays.  Using the average
lepton transverse momenta in $W$ and $Z$ decays as simplified
observables for the determination of the $W$ mass, we estimate that
mixed QCD-electroweak corrections can shift the extracted value of the
$W$ mass by up to ${\cal O}(20)~{\rm MeV}$, depending on the kinematic
cuts employed to define fiducial cross sections for $Z$ and $W$
production.  Since the target precision of the $W$-mass measurement at
the LHC is ${\cal O}(10)~{\rm MeV}$, our results emphasize the need
for fully-differential computations of mixed QCD-electroweak
corrections and a careful analysis of their potential impact on the
determination of the $W$ mass.
 \end{abstract}

\maketitle

\section{Introduction} 
The measurement of the $W$ boson mass at the LHC is a Holy Grail of
precision hadron collider physics. It is believed that the $W$ mass
can be extracted from  LHC data with an uncertainty of about
$10$~{\rm MeV}~\cite{ATLAS:2018qzr}. If this happens, the precision of the direct
measurement  will match the precision that has already been achieved
for the $W$ mass  extracted from global electroweak fits  using the 
renormalizability of the Standard Model~\cite{Baak:2014ora}. A
comparison of direct and indirect determinations of the $W$ mass
has the potential to further stress-test the consistency of the Standard Model at
the quantum level  and, perhaps, reveal unknown contributions to precision electroweak
observables. Until now such a comparison  has been limited by  the
uncertainties  of direct determinations.  Indeed, the $W$ mass
was measured at LEP and Tevatron with an  uncertainty of 33 MeV~\cite{Schael:2013ita}
and 16 MeV~\cite{Aaltonen:2013iut}, respectively. Recently, the ATLAS Collaboration 
reported a measurement  of $m_W$ with  an uncertainty of 19
MeV~\cite{Aaboud:2017svj}.  To improve on this  result, both an
exquisite control of  experimental systematics  and a thorough
investigation of all possible sources of theory uncertainties  are necessary.

In general, measurements of particle masses at  colliders rely on
correlations between them and selected kinematic observables.  A classic
example of such an observable, which has been employed to measure  $m_W$
for many years, is the so-called transverse $W$ mass\footnote{For a definition, see
e.g. Ref.~\cite{Aaboud:2017svj}.},  which has a sharp edge at
$m_W$. The observation of such an edge provides one with immediate
information about the value of the $W$ mass which depends only weakly
on the theoretical description of $W$ production in hadron collisions
and its subsequent decay. Nevertheless, even in this case 
ultra-high precision  on the $W$-mass measurement calls for 
a detailed understanding of  e.g. the uncertainty with which
the missing energy can be determined, the effects of the finite width of the $W$ boson in theoretical modeling
of $W$ production, and so on.

Another important observable that is used for the $W$-mass measurement
is the transverse momentum distribution of charged leptons from the
decay $W \to l \nu_l$.  Features of  this distribution are  correlated with $m_W$ and, in comparison to 
the transverse mass, it is under  better experimental control. As a consequence, the $p_\perp^l$
distribution plays quite a  prominent role in  high-precision
$m_W$ determinations. Indeed, the recent 
ATLAS extraction of the $W$ mass at the LHC \cite{Aaboud:2017svj} was
mostly
driven by the measurement of the charged lepton $p^{l}_\perp$  distribution. 

Unfortunately, the $p_{\perp}^{l}$ distribution is quite sensitive to the
theoretical description of $W$ production and decay, including the modeling
of the transverse momentum spectrum of the $W$ boson, control of the 
parton distribution functions, and a detailed understanding of QCD
\emph{and} QED radiation, both from the initial and the final state.
Although it is well understood how to describe the charged lepton $p_\perp^l$
distribution using the 
framework of collinear factorization in QCD,  the challenge arises from  the extraordinary precision of the  planned
$W$-mass measurement.   Indeed, as
we already mentioned, the $W$ mass is expected to be measured with a 
precision of about ${\cal O}(10)~{\rm MeV}$ or $0.01$ {\it percent}.
It is perfectly clear that existing theoretical approaches, be they
fixed order computations or parton showers or resummations, are not suitable
for the description  of  {\it any} hadron-collider observable with such
precision.

\begin{table*}[t]
\centering
\begin{tabular}{|c|c|c|c||c|c|c|}
\hline
& \multicolumn{3}{|c||}{$V=Z$} & 
\multicolumn{3}{|c|}{$V=W^+$}\\
\cline{1-7}
    & $\mu = m_Z/4$ & $\mu = m_Z/2$ & $\mu = m_Z$  & $\mu=m_W/4$ & $\mu = m_W/2$  & $\mu = m_W$  \\
\hline
\hline 
$F_V(0,0;1),~[\rm{pb}]$ & $1273$ & $1495$ & $1700$ & $7434$ & $8810$ & $10083$ \\
$F_V(1,0;1),~[\rm{pb}]$ & $570.2$ & $405.4$ & $246.9$ & $3502$ & $2533$ &  $1580$ \\
$F_V(0,1;1),~[\rm{pb}]$ & $-5810\cdot 10^{-3}$ & $-6146 \cdot 10^{-3}$ & $-6073\cdot 10^{-3}$ & $-1908 \cdot 10^{-3}$ & $ 3297 \cdot 10^{-3}$ & $10971 \cdot 10^{-3}$ \\
$F_V(1,1;1),~[\rm{pb}]$ & $-2985\cdot 10^{-3}$ & $-2033 \cdot 10^{-3}$ & $-1236\cdot 10^{-3}$ & $-8873 \cdot 10^{-3}$ & $-7607 \cdot 10^{-3}$ & $-7556 \cdot 10^{-3}$ \\
\hline
\hline
$F_V(0,0;p_\perp^e)~[\rm{GeV}\cdot\rm{pb}]$ & $42741$   & $50191$   & $57073$   & $220031$ & $260772$  & $298437$ \\
$F_V(1,0;p_\perp^e)~[\rm{GeV}\cdot\rm{pb}]$ & $23418$   & $17733$   & $12221$   & $124487$ & $95132$   & $66090$  \\
$F_V(0,1;p_\perp^e)~[\rm{GeV}\cdot\rm{pb}]$ & $-182.85$ & $-192.77$ & $-189.11$ & $74.53$  & $243.54$  & $484.82$ \\
$F_V(1,1;p_\perp^e)~[\rm{GeV}\cdot\rm{pb}]$ & $-163.87$ & $-125.22$ & $-92.05$ & $-553.87$ & $-482.0$  & $-448.0$ \\
\hline
\end{tabular}
\caption{Inclusive cross sections and first moments of the positron transverse  momentum distributions  in $pp \to W^+ \to \nu e^+$ and $pp \to Z \to e^- e^+$ at the 13 TeV LHC.
  Results are shown at leading order, for the next-to-leading order  QCD and EW corrections, and for the mixed QCD-electroweak corrections. See text for details.}
\label{resscale}
\end{table*}

This problem is usually overcome by exploiting  similarities between the production
of $Z$ and
$W$ bosons in hadron collisions and by making use of the fact that the mass of the
$Z$ boson has been measured very precisely at LEP.
The extraction of the $W$ mass from studies of the lepton distribution $p_\perp^{l}$
in the process $pp \to W+X \to l \nu_l+ X$ relies on these considerations and 
makes use of the fact that a similar distribution in the process $pp \to Z+X \to l \bar{l}
+X$ can be  used for calibration purposes. The  underlying theoretical
assumption is that QCD effects in these two processes are strongly
correlated and, as a consequence,  a theoretical model  ``tuned''
to describe the $p_\perp^l$ distribution in the $Z$ sample  can be used
with minimal modifications to obtain precise predictions for the $p_\perp^l$
distribution in the $W$ case. This is the approach on which 
the analysis of Ref.~\cite{Aaboud:2017svj} as well as earlier measurements
of the $W$ mass at the Tevatron are based. 

Clearly, if one relies  on using $Z$ boson production for the calibration, all effects that
distinguish between the $Z$ and $W$ cases must be  estimated theoretically. 
 As we already  mentioned, QCD corrections are expected
 to be largely similar for $W$ and $Z$ production,   although   even in this case
 the impact of  different quark flavors in the initial state
 \cite{Bagnaschi:2019mzi,Bozzi:2011ww,Bozzi:2015hha,  Farry:2019rfg, Pietrulewicz:2017gxc,Bagnaschi:2018dnh}  as well as 
of the $gg \to Zg$ contribution  that exists in $Z$ production but not in the $W$ case must be investigated. 

On the other hand,  it is also clear that electroweak (EW) corrections may  affect the production
of $W$ and $Z$ bosons differently,  potentially leading to uncorrelated   effects
of these corrections on the $p_\perp^l$ spectra in $Z$ and $W$ samples. If this does happen,
any measurement of the $W$ mass that relies of the similarity of $Z$ and $W$ kinematic distributions
will be affected.  

These considerations motivated extensive studies of the NLO
electroweak
corrections~\cite{Wackeroth:1996hz,Baur:1997wa,Baur:1998kt,Baur:2001ze,
  Dittmaier:2001ay,Baur:2004ig,Arbuzov:2005dd,Arbuzov:2007db,Dittmaier:2009cr}
to the $Z$ and $W$ production processes, as well as effects related to
multiple photon
emissions~\cite{Barberio:1990ms,Barberio:1993qi,Placzek:2003zg,CarloniCalame:2005vc,Golonka:2005pn,CarloniCalame:2006zq,CarloniCalame:2007cd}
in $Z$ and $W$ decays.  Their impact on the $W$-mass determination has
been studied in detail, see Ref.~\cite{CarloniCalame:2016ouw} for a
comprehensive review.

It was also recognized long ago that for the target precision of the
$W$-mass measurement one has to go beyond NLO electroweak corrections
and account for \emph{mixed} QCD-electroweak effects.  Approximate
$\mathcal{O}(\alpha_s \alpha_W)$ corrections are available in parton
showers using a factorized approach
~\cite{Balossini:2009sa,Bernaciak:2012hj,Barze:2013fru}, and their
impact on the $W$-mass determination was also studied in
Ref.~\cite{CarloniCalame:2016ouw}.  However, genuine mixed QCD-EW
corrections were, until recently, only known for initial-state QCD
radiation and final-state photon
emission~\cite{Dittmaier:2014qza,Dittmaier:2015rxo} which are expected
to give the dominant contribution to the full QCD-EW
corrections. Their impact on $W$-mass determinations has been studied
in Refs.~\cite{Dittmaier:2015rxo,CarloniCalame:2016ouw}.

The computation of the remaining mixed QCD-EW corrections to the $Z$
and $W$ production processes was recently
completed~\cite{deFlorian:2018wcj,Delto:2019ewv,Buccioni:2020cfi,Cieri:2020ikq,Bonciani:2020tvf,Behring:2020cqi,Dittmaier:2020vra,Buonocore:2021rxx}.
The goal of this note is to estimate how these corrections affect the
value of the $W$ mass extracted from the transverse momentum
distribution of a charged lepton.

Although in the experimental analyses ~\cite{Aaltonen:2013iut,Aaboud:2017svj,Cipriani:2019aze}  the mass of the
$W$ boson is determined from fits to templates of $p_\perp^{l}$
distributions, here we adopt a simplified approach that  allows 
us to {\it estimate} the resulting mass  shift in a simple and transparent way. 
We believe that the simplicity and transparency  of our analysis  justifies  its
use in a theoretical paper but we emphasize that, should corrections
turn out to be non-negligible, a more refined study of the  impact of mixed
QCD-EW effects on the $W$-mass extraction that  better reflects the details of 
experimental analyses  will be required. 

To estimate the impact of mixed QCD-electroweak corrections on the $W$-mass
measurement we make use of the fact that the average transverse momentum of the
charged lepton in the Drell-Yan processes $ \langle p_\perp^{l,V} \rangle $ $(V=Z,W)$ 
is correlated with the mass of the respective gauge boson.  Indeed, it is
straightforward to compute $ \langle p_\perp^{l,V} \rangle $ at
leading order in perturbative QCD.
The result, as a function of the lower cut on the lepton transverse momentum
$p_\perp^{\rm cut}$, is
\be
\langle p_\perp^{l,V} \rangle = m_V
f\left ( \frac{p_\perp^{\rm cut}}{M_V} \right ),
\label{eq1a}
\ee
where
\be f\left
( r \right ) = \frac{3}{32} \frac{r (5-8 r^2)}{1-r^2} + \frac{15}{64}
\frac{\arcsin \left ( \sqrt{1-4r^2} \right )}{(1-r^2) \sqrt{1-4r^2}}.
\ee
The function $f(r)$ quantifies  the dependence of the average momentum $ \langle p_\perp^{l,V} \rangle $ 
on  the $p_\perp^{\rm cut}$; if no cut is imposed, we
obtain $\langle p_\perp^{l,V} \rangle = m_V f(0) = 15 \pi/128 \; m_V$.

We note that for physical values of $r$,  $ 0 < r < 0.5$,  the function $f(r)$ does not change strongly, $0.368 < f(r) < 0.5$.
Therefore, we expect that
either the selection of cuts can be optimized to enhance the similarity of the $p_\perp^l$  distributions
in $W$ and $Z$ production, or that the effect of cuts can be adequately predicted in
perturbation theory.  
Hence, we write the following formula for the
$W$ mass extracted from measurements of  average values of lepton transverse momenta as
\be
m_W^{\rm meas} = \frac{ \langle p_\perp^{l,W} \rangle^{\rm meas} }{ \langle p_\perp^{l,Z} \rangle^{\rm meas } }\; m_Z \; C_{\rm th}.
\label{eq1}
\ee
The {\it theoretical} correction factor $C_{\rm th}$ is determined by
comparing the value of the $W$ mass obtained by following this procedure
within a particular theoretical framework with the actual $W$ mass
$m_W$ used as an input in a theoretical calculation.  Therefore
\be
C_{\rm th} = \frac{m_W}{m_Z} \frac{ \langle p_\perp^{l,Z}  \rangle^{\rm th} }{ \langle p_\perp^{l,W} \rangle^{\rm th } }.
\label{eq2}
\ee

If the theoretical framework used to compute $C_{\rm th}$ changes, for
example because a more refined  theoretical prediction
for $ \langle p_\perp^{l} \rangle $
becomes available, there is a shift in the
extracted value of the $W$ mass $m_W^{\rm meas}$.  It evaluates to
\be
\frac{\delta m_W^{\rm meas}}{m_W^{\rm meas}} = \frac{ \delta C_{\rm th}}{C_{\rm th}} = \frac{ \delta
  \langle p_\perp^{l,Z}  \rangle^{\rm th} }{ \langle p_\perp^{l,Z} \rangle^{\rm th } }
-
\frac{ \delta
  \langle p_\perp^{l,W}  \rangle^{\rm th} }{ \langle p_\perp^{l,W} \rangle^{\rm th } }.
\label{eq3}
\ee
This equation  shows clearly the role that the $Z$ boson observables
play in Eqs.(\ref{eq1},\ref{eq2}).  Indeed, it follows from
Eq.(\ref{eq3}) that all effects that influence the lepton transverse momentum
distributions in
$Z$ and $W$ production and decay in a similar way do not result in
a shift in the measured value of the $W$ mass.  However, if this is not
the case, a shift in the extracted value $m_W^{\rm meas}$ arises.

Eq.(\ref{eq3}) provides the  basis for our  estimate of
the impact of the mixed QCD-electroweak corrections on the determination
of the $W$ mass. Indeed, the calculations reported in
Refs.~\cite{Buccioni:2020cfi,Behring:2020cqi} allow us to compute
average lepton transverse momenta  in $Z$ and $W$ production with and
without mixed QCD-electroweak corrections. Using this information, we
construct quantities that appear on  the right hand side  of Eq.(\ref{eq3}) and
estimate the shift in the extracted value of the $W$ mass.

Before presenting the  results, we briefly discuss the setup of the calculation.  We use
the same input parameters as described in
Refs.~\cite{Buccioni:2020cfi,Behring:2020cqi}. In particular, we adopt
the $G_{\mu}$ renormalization scheme and use $G_F=1.16639\cdot 10^{-5}~\rm{GeV}^{-2}$,
$m_Z=91.1876~\rm{GeV}$, $m_W=80.398~\rm{GeV}$, $m_H=125~\rm{GeV}$ and 
$m_t=173.2~\rm{GeV}$. We work in the narrow-width approximation
and consider all quarks but the top quark  to be massless.\footnote{We neglect the contribution
  of Feynman diagrams with internal top quarks in the calculation of
  mixed QCD-electroweak two-loop corrections
  Our result then only
  depends on $m_t$ through the renormalization procedure, see
  Ref.~\cite{Buccioni:2020cfi} for details.}
For definiteness, we consider decays $Z \to e^-e^+$ and $W^+ \to \nu_e e^+$ and consider the electrons as being massless.
We employ 
the NNLO
NNPDF3.1luxQED~\cite{Bertone:2017bme,Manohar:2017eqh,Manohar:2016nzj}
parton distributions with  $\alpha_s(m_Z)=0.118$.
For our analysis, we
focus on $Z$ and $W^+$ production at the 13 TeV LHC and study the transverse
momentum distribution of the positron $e^+$.  Since the
contribution of QCD initial-state and EW final-state corrections to the
full mixed QCD-EW result and its impact on the $W$-mass determinations is 
known~\cite{Dittmaier:2014qza,Dittmaier:2015rxo},
we do not consider corrections to the  $W \to \nu_e e^+$ and $Z\to e^-e^+$ decay subprocesses.
In other words, for our estimates we only consider mixed QCD-EW
corrections to the {\it production} sub-processes $pp\to W/Z$. As we have already
said, this is the only mixed QCD-electroweak  contribution  whose impact on the $W$-mass determination is
currently unknown. 

For the sake of clarity, we begin by considering inclusive quantities
and do not apply any kinematic cuts.
We write the differential cross sections for $Z$ and $W$
production as
\be
   {\rm d} \sigma_{Z,W} = \sum \limits_{i,j=0} \alpha_s^{i} \alpha_W^{i} {\rm d} \sigma^{i,j}_{Z,W},  
   \ee
   where $\alpha_s$ and $\alpha_W$ are the strong and electroweak couplings, respectively. We also define weighted integrals 
\be
F_{Z,W}(i,j,{\cal O}) =  \alpha_s^{i} \alpha_W^{i} \int {\rm d} \sigma^{i,j}_{Z,W}  \times {\cal O},
\ee
where ${\cal O}$ is a particular kinematic variable.  With this notation, 
the average transverse momentum of the positron in the processes
$pp \to Z +X \to e^- e^+ + X$ and $pp \to W^+ +X \to \nu_e e^+ + X$ reads
\be
\langle p_{\perp}^{e^+,V} \rangle^{\rm th}
= \frac{\sum \limits_{ij} F_{V}( i, j,p_\perp^{e^+}) }{
\sum \limits_{ij} F_{V}( i, j,1)
}.
\label{eq5}
\ee
In Table~\ref{resscale}  we report results for $F_{V}$  when no  fiducial cuts are applied.

To study the impact of mixed QCD-EW corrections on the $W$-mass
determination, we use Eq.\eqref{eq3}. We
determine the shifts $\delta \langle p_\perp^{e^+,V} \rangle^{\rm th}$
by computing $ \langle p_{\perp}^{e^+,V}
\rangle^{\rm th} $ in Eq.(\ref{eq5}) with mixed
QCD-electroweak contributions (i.e. with the $F_V(1,1,\dots)$
terms). We then take the difference of this result with respect to the result including both the NLO QCD and  NLO EW corrections.
Using the results presented in
Table~\ref{resscale}, we find 
\be
\frac{\delta m_W^{\rm meas}}{m^{\rm meas}_W } =
-0.93^{-0.22}_{+0.29} \times 10^{-4}.
\label{eq7}
\ee
To compute  the central value, we have set both the renormalization and
factorization scales to $\mu = m_V/2$. The upper (lower)
value corresponds to $\mu = m_V$ and $\mu = m_V/4$, respectively.

Using $m_W^{\rm meas} = 80.398~{\rm GeV}$ in Eq.(\ref{eq7}), we
find that the value of the $W$ boson mass extracted
from the  $\langle p_\perp^{e^+} \rangle$  distribution without 
accounting for mixed QCD-electroweak corrections exceeds the true
value by ${\cal O}(7)~{\rm MeV}$. This result is only mildly affected
by PDFs uncertainties: using a compressed NNPDF3.1luxQED set, obtained along the lines described in
  Refs.~\cite{Carrazza:2016htc, Carrazza:2016wte},
we find that  uncertainties in parton distribution
functions may change 
the above estimate of the mass shift by   about 1 MeV.

It is interesting to point out that if we use this analysis to study
the impact of {\it electroweak} corrections to the production processes
$pp \to Z$ and $pp \to W^+$  on the value of the $W$ mass, we find a very small shift
of about ${\cal O}(1)~{\rm MeV}$  provided that  we use the NLO QCD calculation as a
baseline. This result shows  that mixed
QCD-electroweak corrections have {\it larger impact} on the $W$-mass
measurement than the electroweak ones.
There seem to be two reasons for that. The first reason is that electroweak
and mixed QCD-electroweak corrections to observables in $W$ and $Z$
production are comparable and  do not quite follow the standard
hierarchy where the electroweak corrections are expected to be larger than
the mixed ones.
This feature can be seen in Table~\ref{resscale}, and was also previously noted in
Refs.~\cite{Buccioni:2020cfi,Behring:2020cqi} where it was pointed out
that the use of the so-called $G_\mu$ renormalization scheme
 reduces electroweak corrections significantly.  The second reason for the
tiny shift in the extracted value of the $W$ mass caused by the electroweak
corrections is a very strong cancellation between the first and the second
terms on the right hand side  of Eq.(\ref{eq3}).  This means that electroweak 
corrections cause nearly identical relative changes in the average transverse
momenta of charged leptons in decays of $Z$ and $W$
bosons, so that the significance of these corrections is substantially reduced. 

To elaborate on  this point further, we note that if we only compute
relative changes to the average transverse momentum of the lepton coming from the $W$ decay
and set the term
$ \delta \langle p_{\perp}^{e^+,Z} \rangle / \langle p_{\perp}^{e^+,Z} \rangle $
in Eq.(\ref{eq3}) to zero, we find that electroweak corrections
induce a ${\cal  O}(-31)~{\rm MeV}$ shift in $m_W$.  If we do the same for mixed
QCD-electroweak corrections, this mass shift turns out to be ${\cal
  O}(54)~{\rm MeV}$.     These results imply that  {\it i}) the magnitude of EW
and QCD-EW corrections to the average lepton transverse momenta  are
indeed comparable; {\it ii}) there are significant correlations between
corrections to average $p_\perp^{e^+}$ in $Z$ and $W$
production and {\it iii}) these correlations are {\it slightly}  stronger for
electroweak than for mixed QCD-electroweak corrections leading to {\it significantly}
larger  shifts in $m_W^{\rm meas}$ in the latter case. 

We can easily extend the calculation  that we just described to include
kinematic restrictions applied  in experimental analyses. As an example,
we re-compute the  average  transverse momenta of the charged leptons
using kinematic cuts inspired by the ATLAS analysis~\cite{Aaboud:2017svj}. In the case of  $W$ boson
production, we require that the transverse momentum of the charged
lepton and the missing transverse momentum, which we identify with the
transverse momentum of the neutrino, satisfy $p_{\perp}^{e^+} > 30~{\rm  GeV}$
and $p_\perp^{\rm miss} > 30~{\rm GeV}$, and that the rapidity of
the charged lepton is bounded by $|\eta_{e^+}| < 2.4$. We also require that the
transverse mass of the positron-neutrino system is larger than
$60~{\rm GeV}$.  In the case of the $Z$ boson, we select electrons and
positrons with transverse momenta larger than $25~{\rm GeV}$ and
require that their rapidities are within the interval $|\eta_{e^\pm}| < 2.4$.

Repeating the  computation described above for fiducial cross sections,
we find larger shifts in the $W$ mass due to mixed QCD-electroweak corrections.
Specifically, we obtain
\be
\delta m_W^{\rm meas} = -17\pm 2 ~{\rm MeV},
\label{eq10a}
\ee
where the central value is for $\mu = m_V/2$ and the uncertainty
is obtained from a three-point scale variation.  Although electroweak corrections
also increase if fiducial cuts are applied, they are  still small; we estimate that they
change  the measured value of the $W$ mass by only about $3~{\rm MeV}$.

Although a detailed study of the impact of fiducial cuts on
the $W$-mass extraction  is beyond the scope of this simple
analysis, it is interesting to investigate how the somewhat larger
$\mathcal O(17)~{\rm MeV}$ shift comes about.  The key reason for this 
is that the transverse momenta  that play  a role in the
analysis  are determined by  ratios  $p^{e^+}_\perp/M_V$, see Eq.(\ref{eq1a}).
The ATLAS collaboration applies a {\it  higher}  $p^{e^+}_\perp$ cut to  the  (lighter)  $W$ boson sample than to  the (heavier) Z boson sample. 
Effectively,  this choice of cuts  gives  higher weight
to the high-$p_\perp^{e^+}$ region in the $W$ case
as compared to  the $Z$ case.
Since radiative corrections in the $W$ case extend to a wider range beyond
the Jacobian peak, this  leads to a (small) decorrelation of the transverse momentum distributions
from $Z$ and $W$ production~\cite{Giele:1998uh} which is  sufficient, however, to cause 
a shift in $m_W$  that appears to be significant  given the target precision. 

If the large shift in the $W$ mass in Eq.(\ref{eq10a}) is caused by an experimentally motivated but
``unfortunate'' choice of cuts, one can ask whether it possible to choose cuts in such a way that the
similarity of $p^{e^+}_\perp$-distributions in $Z$ and $W$ samples is actually enforced.
To answer this question, we proceed as follows:
we start with the baseline fiducial region described above, but for
the $W^+$ case we \emph{decrease} the cuts on the transverse momentum of the positron
and on the missing transverse energy until 
the  theoretical  correction  factor $C_{\rm th}$ in 
Eq.~\eqref{eq2} becomes $C_{\rm th} = 1$ at leading order.
This leads to a $p_\perp$ cut of $25.44~{\rm GeV}$. 
Using this set of  cuts, we find that both the EW and the mixed QCD-EW corrections to the
$p_{\perp}^{e^+}$ spectra in $Z$ and $W$ production become more strongly correlated.  Specifically, we observe that
mixed QCD-EW corrections shift the $W$ mass by only 
\be
\delta m_W^{\rm meas} = -1 \pm 5~{\rm MeV},
\ee
where again the central value corresponds to $\mu = m_V/2$ and the
uncertainty is obtained from a three-point scale variation. For comparison,
electroweak corrections in this case shift the $W$ mass by  $\mathcal O(-3)~{\rm MeV}$. 

In conclusion,  we have applied a simple and
theoretically clean procedure to estimate the impact of the recently
computed mixed QCD-EW
corrections~\cite{Buccioni:2020cfi,Behring:2020cqi} on the $W$-mass
extraction  at the LHC. Similar to the experimental
analyses~\cite{Aaltonen:2013iut,Aaboud:2017svj,Cipriani:2019aze},
we used the transverse momentum distribution of a charged lepton from $W$ decays 
as an observable from which the $W$ mass can be inferred. 
However, instead of using the full distribution, 
we focused our analysis on its first moment, i.e.  on the average  $p_\perp$ of the charged lepton.

The key element of the experimental analysis is the use of the lepton $p_\perp$ distribution
in $Z$ production and the known mass of the $Z$ boson as a constraint  to be employed in the extraction of the $W$ mass. 
The idea is that all effects  that impact  $Z$ and $W$ production  in a similar way play
no role in the $W$-mass extraction
if the $Z$ sample is used to normalize the $W$ sample.  Hence, the important question is not
by how much lepton distributions in $Z$ and $W$ production are affected by various radiative corrections but rather
if they are affected in a correlated fashion or not.

Our calculations and analyses indicate that there is no simple answer to this question in the sense that
selection criteria applied to $Z$ and $W$ samples do  matter~\cite{Giele:1998uh}.  Indeed, we observe that when no cuts are applied 
to lepton $p_\perp$ distributions or when the $p_\perp^{l,W, Z}$ cuts are chosen in a way that roughly respect the
ratio of $W$ and $Z$ masses,  shifts in $m_W$ caused by the mixed QCD-electroweak corrections to the production
process appear to be below the LHC target precision of ${\cal O}(10)$~{\rm MeV}. On the other hand,
with a choice of cuts more aligned with experimental practices we find that
mixed QCD-electroweak corrections
cause bigger shifts in $m_W$.  For example, we estimate that the cuts employed by the ATLAS collaboration in their recent
extraction of the $W$ mass~\cite{Aaboud:2017svj} may lead  to a shift of about ${\cal O}(17)$~MeV due to
unaccounted mixed QCD-electroweak effects in the production process.

We stress that these results
are only \emph{estimates}: given the simplified nature of our analysis,
we cannot insist that shifts in the $W$ mass  described above  should be applied to results
of actual  measurements.  
Nevertheless, we believe that the size of the effects found
here warrants  further, more in-depth studies, which
should ideally go hand-in-hand with the actual
experimental analyses. Natural avenues of investigation include using
more differential information rather than just the first moment of
the lepton $p_\perp$ distribution, or quantifying how much of these effects are actually
captured by the simulation tools that are currently used by the
experimental collaborations. We believe that such studies are mandatory
to make a convincing case for  $\mathcal O(10)~{\rm MeV}$ precision on $W$-mass
extractions at the LHC.

{\bf Acknowledgment} We would like to thank the NNPDF collaboration
and especially S.  Carrazza for providing the compressed PDF set that
we used in our analysis. We are grateful to  M.~Boonekamp
and G. Salam for interesting discussions. This research is partially
supported by the Deutsche Forschungsgemeinschaft (DFG, German Research
Foundation) under grant 396021762 - TRR 257. The research of F.B. and
F.C. was partially supported by the ERC Starting Grant 804394 HipQCD.


\begin{thebibliography}{99}

%\cite{ATLAS:2018qzr}
\bibitem{ATLAS:2018qzr}
 [ATLAS],
%``Prospects for the measurement of the W-boson mass at the HL- and HE-LHC,''
ATL-PHYS-PUB-2018-026.
%15 citations counted in INSPIRE as of 22 Feb 2021
  
%\cite{Baak:2014ora}
\bibitem{Baak:2014ora}
M.~Baak \textit{et al.} [Gfitter Group],
%``The global electroweak fit at NNLO and prospects for the LHC and ILC,''
Eur. Phys. J. C \textbf{74}, 3046 (2014).
%doi:10.1140/epjc/s10052-014-3046-5
%[arXiv:1407.3792 [hep-ph]].

%\cite{Schael:2013ita}
\bibitem{Schael:2013ita} S.~Schael \textit{et al.} [ALEPH, DELPHI, L3,
  OPAL and LEP Electroweak],
%``Electroweak Measurements in Electron-Positron Collisions at
%W-Boson-Pair Energies at LEP,''
Phys. Rept. \textbf{532}, 119-244 (2013).
%doi:10.1016/j.physrep.2013.07.004
%[arXiv:1302.3415 [hep-ex]].
%541 citations counted in INSPIRE as of 22 Feb 2021

%\cite{Aaltonen:2013iut}
\bibitem{Aaltonen:2013iut}
T.~A.~Aaltonen \textit{et al.} [CDF and D0],
%``Combination of CDF and D0 $W$-Boson Mass Measurements,''
Phys. Rev. D \textbf{88}, no.5, 052018 (2013).
%doi:10.1103/PhysRevD.88.052018
%[arXiv:1307.7627 [hep-ex]].
%116 citations counted in INSPIRE as of 22 Feb 2021

%\cite{Aaboud:2017svj}
\bibitem{Aaboud:2017svj}
M.~Aaboud \textit{et al.} [ATLAS],
%``Measurement of the $W$-boson mass in pp collisions at $\sqrt{s}=7$ TeV with the ATLAS detector,''
Eur. Phys. J. C \textbf{78}, no.2, 110 (2018)
[erratum: Eur. Phys. J. C \textbf{78}, no.11, 898 (2018)].
%doi:10.1140/epjc/s10052-017-5475-4
%[arXiv:1701.07240 [hep-ex]].
%238 citations counted in INSPIRE as of 22 Feb 2021

%\cite{Bagnaschi:2019mzi}
\bibitem{Bagnaschi:2019mzi}
E.~Bagnaschi and A.~Vicini,
%``Parton Density Uncertainties and the Determination of Electroweak Parameters at Hadron Colliders,''
Phys. Rev. Lett. \textbf{126}, no.4, 041801 (2021).
%doi:10.1103/PhysRevLett.126.041801
%[arXiv:1910.04726 [hep-ph]].
%6 citations counted in INSPIRE as of 22 Feb 2021

%\cite{Bozzi:2011ww}
\bibitem{Bozzi:2011ww}
G.~Bozzi, J.~Rojo and A.~Vicini,
%``The Impact of PDF uncertainties on the measurement of the W boson
%mass at the Tevatron and the LHC,''
Phys. Rev. D \textbf{83}, 113008 (2011).
%doi:10.1103/PhysRevD.83.113008
%[arXiv:1104.2056 [hep-ph]].
%59 citations counted in INSPIRE as of 22 Feb 2021

%\cite{Bozzi:2015hha}
\bibitem{Bozzi:2015hha}
G.~Bozzi, L.~Citelli and A.~Vicini,
%``Parton density function uncertainties on the W boson mass measurement from the lepton transverse momentum distribution,''
Phys. Rev. D \textbf{91}, no.11, 113005 (2015).
%doi:10.1103/PhysRevD.91.113005
%[arXiv:1501.05587 [hep-ph]].
%32 citations counted in INSPIRE as of 22 Feb 2021

%\cite{Farry:2019rfg}
\bibitem{Farry:2019rfg}
S.~Farry, O.~Lupton, M.~Pili and M.~Vesterinen,
%``Understanding and constraining the PDF uncertainties in a $W$ boson mass measurement with forward muons at the LHC,''
Eur. Phys. J. C \textbf{79}, no.6, 497 (2019).
%doi:10.1140/epjc/s10052-019-6997-8
%[arXiv:1902.04323 [hep-ex]].
%7 citations counted in INSPIRE as of 22 Feb 2021

%\cite{Pietrulewicz:2017gxc}
\bibitem{Pietrulewicz:2017gxc}
P.~Pietrulewicz, D.~Samitz, A.~Spiering and F.~J.~Tackmann,
%``Factorization and Resummation for Massive Quark Effects in Exclusive Drell-Yan,''
JHEP \textbf{08}, 114 (2017).
%doi:10.1007/JHEP08(2017)114
%[arXiv:1703.09702 [hep-ph]].
%12 citations counted in INSPIRE as of 22 Feb 2021

%\cite{Bagnaschi:2018dnh}
\bibitem{Bagnaschi:2018dnh}
E.~Bagnaschi, F.~Maltoni, A.~Vicini and M.~Zaro,
%``Lepton-pair production in association with a $ b\overline{b} $ pair and the determination of the $W$ boson mass,''
JHEP \textbf{07}, 101 (2018).
%doi:10.1007/JHEP07(2018)101
%[arXiv:1803.04336 [hep-ph]].
%13 citations counted in INSPIRE as of 22 Feb 2021

%\cite{Wackeroth:1996hz}
\bibitem{Wackeroth:1996hz}
D.~Wackeroth and W.~Hollik,
%``Electroweak radiative corrections to resonant charged gauge boson production,''
Phys. Rev. D \textbf{55}, 6788-6818 (1997).
%doi:10.1103/PhysRevD.55.6788
%[arXiv:hep-ph/9606398 [hep-ph]].
%93 citations counted in INSPIRE as of 22 Feb 2021

%\cite{Baur:1997wa}
\bibitem{Baur:1997wa}
U.~Baur, S.~Keller and W.~K.~Sakumoto,
%``QED radiative corrections to $Z$ boson production and the forward backward asymmetry at hadron colliders,''
Phys. Rev. D \textbf{57}, 199-215 (1998).
%doi:10.1103/PhysRevD.57.199
%[arXiv:hep-ph/9707301 [hep-ph]].
%157 citations counted in INSPIRE as of 22 Feb 2021

%\cite{Baur:1998kt}
\bibitem{Baur:1998kt}
U.~Baur, S.~Keller and D.~Wackeroth,
%``Electroweak radiative corrections to $W$ boson production in hadronic collisions,''
Phys. Rev. D \textbf{59}, 013002 (1999).
%doi:10.1103/PhysRevD.59.013002
%[arXiv:hep-ph/9807417 [hep-ph]].
%222 citations counted in INSPIRE as of 22 Feb 2021

%\cite{Baur:2001ze}
\bibitem{Baur:2001ze}
U.~Baur, O.~Brein, W.~Hollik, C.~Schappacher and D.~Wackeroth,
%``Electroweak radiative corrections to neutral current Drell-Yan processes at hadron colliders,''
Phys. Rev. D \textbf{65}, 033007 (2002).
%doi:10.1103/PhysRevD.65.033007
%[arXiv:hep-ph/0108274 [hep-ph]].
%259 citations counted in INSPIRE as of 22 Feb 2021

%\cite{Dittmaier:2001ay}
\bibitem{Dittmaier:2001ay}
S.~Dittmaier and M.~Kr\"amer,
%``Electroweak radiative corrections to W boson production at hadron colliders,''
Phys. Rev. D \textbf{65}, 073007 (2002).
%doi:10.1103/PhysRevD.65.073007
%[arXiv:hep-ph/0109062 [hep-ph]].
%238 citations counted in INSPIRE as of 22 Feb 2021

%\cite{Baur:2004ig}
\bibitem{Baur:2004ig}
U.~Baur and D.~Wackeroth,
%``Electroweak radiative corrections to $p \bar{p} \to W^\pm \to \ell^\pm \nu$ beyond the pole approximation,''
Phys. Rev. D \textbf{70}, 073015 (2004).
%doi:10.1103/PhysRevD.70.073015
%[arXiv:hep-ph/0405191 [hep-ph]].
%119 citations counted in INSPIRE as of 22 Feb 2021

%\cite{Arbuzov:2005dd}
\bibitem{Arbuzov:2005dd}
A.~Arbuzov, D.~Bardin, S.~Bondarenko, P.~Christova, L.~Kalinovskaya, G.~Nanava and R.~Sadykov,
%``One-loop corrections to the Drell-Yan process in SANC. I. The Charged current case,''
Eur. Phys. J. C \textbf{46}, 407-412 (2006)
[erratum: Eur. Phys. J. C \textbf{50}, 505 (2007)].
%doi:10.1140/epjc/s2006-02505-y
%[arXiv:hep-ph/0506110 [hep-ph]].
%98 citations counted in INSPIRE as of 22 Feb 2021

%\cite{Arbuzov:2007db}
\bibitem{Arbuzov:2007db}
A.~Arbuzov, D.~Bardin, S.~Bondarenko, P.~Christova, L.~Kalinovskaya, G.~Nanava and R.~Sadykov,
%``One-loop corrections to the Drell--Yan process in SANC. (II). The Neutral current case,''
Eur. Phys. J. C \textbf{54}, 451-460 (2008).
%doi:10.1140/epjc/s10052-008-0531-8
%[arXiv:0711.0625 [hep-ph]].
%90 citations counted in INSPIRE as of 22 Feb 2021

%\cite{Dittmaier:2009cr}
\bibitem{Dittmaier:2009cr}
S.~Dittmaier and M.~Huber,
%``Radiative corrections to the neutral-current Drell-Yan process in the Standard Model and its minimal supersymmetric extension,''
JHEP \textbf{01}, 060 (2010).
%doi:10.1007/JHEP01(2010)060
%[arXiv:0911.2329 [hep-ph]].
%146 citations counted in INSPIRE as of 22 Feb 2021




%\cite{Barberio:1990ms}
\bibitem{Barberio:1990ms}
E.~Barberio, B.~van Eijk and Z.~Was,
%``PHOTOS: A Universal Monte Carlo for QED radiative corrections in decays,''
Comput. Phys. Commun. \textbf{66}, 115-128 (1991).
%doi:10.1016/0010-4655(91)90012-A
%466 citations counted in INSPIRE as of 22 Feb 2021

%\cite{Barberio:1993qi}
\bibitem{Barberio:1993qi}
E.~Barberio and Z.~Was,
%``PHOTOS: A Universal Monte Carlo for QED radiative corrections. Version 2.0,''
Comput. Phys. Commun. \textbf{79}, 291-308 (1994).
%doi:10.1016/0010-4655(94)90074-4
%909 citations counted in INSPIRE as of 22 Feb 2021

%\cite{Placzek:2003zg}
\bibitem{Placzek:2003zg}
W.~Placzek and S.~Jadach,
%``Multiphoton radiation in leptonic W boson decays,''
Eur. Phys. J. C \textbf{29}, 325-339 (2003).
%doi:10.1140/epjc/s2003-01223-4
%[arXiv:hep-ph/0302065 [hep-ph]].
%97 citations counted in INSPIRE as of 22 Feb 2021

%\cite{CarloniCalame:2005vc}
\bibitem{CarloniCalame:2005vc}
C.~M.~Carloni Calame, G.~Montagna, O.~Nicrosini and M.~Treccani,
%``Multiple photon corrections to the neutral-current Drell-Yan process,''
JHEP \textbf{05}, 019 (2005).
%doi:10.1088/1126-6708/2005/05/019
%[arXiv:hep-ph/0502218 [hep-ph]].
%76 citations counted in INSPIRE as of 22 Feb 2021

%\cite{Golonka:2005pn}
\bibitem{Golonka:2005pn}
P.~Golonka and Z.~Was,
%``PHOTOS Monte Carlo: A Precision tool for QED corrections in $Z$ and $W$ decays,''
Eur. Phys. J. C \textbf{45}, 97-107 (2006).
%doi:10.1140/epjc/s2005-02396-4
%[arXiv:hep-ph/0506026 [hep-ph]].
%1228 citations counted in INSPIRE as of 22 Feb 2021

\bibitem{CarloniCalame:2006zq}
C.~M.~Carloni Calame, G.~Montagna, O.~Nicrosini and A.~Vicini,
%``Precision electroweak calculation of the charged current Drell-Yan process,''
JHEP \textbf{12}, 016 (2006).
%doi:10.1088/1126-6708/2006/12/016
%[arXiv:hep-ph/0609170 [hep-ph]].
%153 citations counted in INSPIRE as of 22 Feb 2021

%\cite{CarloniCalame:2007cd}
\bibitem{CarloniCalame:2007cd}
C.~M.~Carloni Calame, G.~Montagna, O.~Nicrosini and A.~Vicini,
%``Precision electroweak calculation of the production of a high transverse-momentum lepton pair at hadron colliders,''
JHEP \textbf{10}, 109 (2007).
%doi:10.1088/1126-6708/2007/10/109
%[arXiv:0710.1722 [hep-ph]].
%230 citations counted in INSPIRE as of 22 Feb 2021

%\cite{CarloniCalame:2016ouw}
\bibitem{CarloniCalame:2016ouw}
C.~M.~Carloni Calame, M.~Chiesa, H.~Martinez, G.~Montagna, O.~Nicrosini, F.~Piccinini and A.~Vicini,
%``Precision Measurement of the W-Boson Mass: Theoretical Contributions and Uncertainties,''
Phys. Rev. D \textbf{96}, no.9, 093005 (2017).
%doi:10.1103/PhysRevD.96.093005
%[arXiv:1612.02841 [hep-ph]].
%35 citations counted in INSPIRE as of 22 Feb 2021

%\cite{Balossini:2009sa}
\bibitem{Balossini:2009sa}
G.~Balossini, G.~Montagna, C.~M.~Carloni Calame, M.~Moretti, O.~Nicrosini, F.~Piccinini, M.~Treccani and A.~Vicini,
%``Combination of electroweak and QCD corrections to single W production at the Fermilab Tevatron and the CERN LHC,''
JHEP \textbf{01}, 013 (2010).
%doi:10.1007/JHEP01(2010)013
%[arXiv:0907.0276 [hep-ph]].
%59 citations counted in INSPIRE as of 01 Mar 2021

%\cite{Bernaciak:2012hj}
\bibitem{Bernaciak:2012hj}
C.~Bernaciak and D.~Wackeroth,
%``Combining NLO QCD and Electroweak Radiative Corrections to W boson Production at Hadron Colliders in the POWHEG Framework,''
Phys. Rev. D \textbf{85}, 093003 (2012).
%doi:10.1103/PhysRevD.85.093003
%[arXiv:1201.4804 [hep-ph]].
%60 citations counted in INSPIRE as of 02 Mar 2021

%\cite{Barze:2013fru}
\bibitem{Barze:2013fru}
L.~Barze, G.~Montagna, P.~Nason, O.~Nicrosini, F.~Piccinini and A.~Vicini,
%``Neutral current Drell-Yan with combined QCD and electroweak corrections in the POWHEG BOX,''
Eur. Phys. J. C \textbf{73}, no.6, 2474 (2013).
%doi:10.1140/epjc/s10052-013-2474-y
%[arXiv:1302.4606 [hep-ph]].
%100 citations counted in INSPIRE as of 01 Mar 2021

%\cite{Dittmaier:2014qza}
\bibitem{Dittmaier:2014qza}
S.~Dittmaier, A.~Huss and C.~Schwinn,
%``Mixed QCD-electroweak $\mathcal{O}(\alpha_s\alpha)$ corrections to Drell-Yan processes in the resonance region: pole approximation and non-factorizable corrections,''
Nucl. Phys. B \textbf{885}, 318-372 (2014).
%doi:10.1016/j.nuclphysb.2014.05.027
%[arXiv:1403.3216 [hep-ph]].
%93 citations counted in INSPIRE as of 22 Feb 2021

%\cite{Dittmaier:2015rxo}
\bibitem{Dittmaier:2015rxo}
S.~Dittmaier, A.~Huss and C.~Schwinn,
%``Dominant mixed QCD-electroweak O($\alpha_s\alpha$) corrections to Drell\textendash{}Yan processes in the resonance region,''
Nucl. Phys. B \textbf{904}, 216-252 (2016).
%doi:10.1016/j.nuclphysb.2016.01.006
%[arXiv:1511.08016 [hep-ph]].
%65 citations counted in INSPIRE as of 22 Feb 2021

%\cite{deFlorian:2018wcj}
\bibitem{deFlorian:2018wcj}
D.~de Florian, M.~Der and I.~Fabre,
%``QCD$\oplus$QED NNLO corrections to Drell Yan production,''
Phys. Rev. D \textbf{98}, no.9, 094008 (2018).
%doi:10.1103/PhysRevD.98.094008
%[arXiv:1805.12214 [hep-ph]].
%36 citations counted in INSPIRE as of 01 Mar 2021



%\cite{Delto:2019ewv}
\bibitem{Delto:2019ewv}
M.~Delto, M.~Jaquier, K.~Melnikov and R.~R\"ontsch,
%``Mixed QCD$\otimes$QED corrections to on-shell $Z$ boson production at the LHC,''
JHEP \textbf{01}, 043 (2020).
%doi:10.1007/JHEP01(2020)043
%[arXiv:1909.08428 [hep-ph]].
%24 citations counted in INSPIRE as of 01 Mar 2021

%\cite{Buccioni:2020cfi}
\bibitem{Buccioni:2020cfi}
F.~Buccioni, F.~Caola, M.~Delto, M.~Jaquier, K.~Melnikov and R.~R\"ontsch,
%``Mixed QCD-electroweak corrections to on-shell Z production at the LHC,''
Phys. Lett. B \textbf{811}, 135969 (2020).
%doi:10.1016/j.physletb.2020.135969
%[arXiv:2005.10221 [hep-ph]].
%10 citations counted in INSPIRE as of 22 Feb 2021

%\cite{Cieri:2020ikq}
\bibitem{Cieri:2020ikq}
L.~Cieri, D.~de Florian, M.~Der and J.~Mazzitelli,
%``Mixed QCD\ensuremath{\otimes}QED corrections to exclusive Drell Yan production using the q$_{T}$ -subtraction method,''
JHEP \textbf{09}, 155 (2020).
%doi:10.1007/JHEP09(2020)155
%[arXiv:2005.01315 [hep-ph]].
%11 citations counted in INSPIRE as of 01 Mar 2021

%\cite{Bonciani:2020tvf}
\bibitem{Bonciani:2020tvf}
R.~Bonciani, F.~Buccioni, N.~Rana and A.~Vicini,
%``Next-to-Next-to-Leading Order Mixed QCD-Electroweak Corrections to on-Shell Z Production,''
Phys. Rev. Lett. \textbf{125}, no.23, 232004 (2020).
%doi:10.1103/PhysRevLett.125.232004.
%[arXiv:2007.06518 [hep-ph]].
%5 citations counted in INSPIRE as of 22 Feb 2021



%\cite{Behring:2020cqi}
\bibitem{Behring:2020cqi}
A.~Behring, F.~Buccioni, F.~Caola, M.~Delto, M.~Jaquier, K.~Melnikov and R.~R\"ontsch,
%``Mixed QCD-electroweak corrections to $W$-boson production in hadron collisions,''
Phys. Rev. D \textbf{103}, no.1, 013008 (2021).
%doi:10.1103/PhysRevD.103.013008
%[arXiv:2009.10386 [hep-ph]].
%2 citations counted in INSPIRE as of 22 Feb 2021

%\cite{Dittmaier:2020vra}
\bibitem{Dittmaier:2020vra}
S.~Dittmaier, T.~Schmidt and J.~Schwarz,
%``Mixed NNLO QCD\texttimes{}electroweak corrections of $\mathcal{O}(N_f \alpha_s \alpha)$ to single-W/Z production at the LHC,''
JHEP \textbf{12}, 201 (2020).
%doi:10.1007/JHEP12(2020)201
%[arXiv:2009.02229 [hep-ph]].
%4 citations counted in INSPIRE as of 01 Mar 2021

%\cite{Buonocore:2021rxx}
\bibitem{Buonocore:2021rxx}
L.~Buonocore, M.~Grazzini, S.~Kallweit, C.~Savoini and F.~Tramontano,
%``Mixed QCD-EW corrections to $\boldsymbol{pp\!\to\!\ell\nu_\ell\!+\!X}$ at the LHC,''
[arXiv:2102.12539 [hep-ph]].
%0 citations counted in INSPIRE as of 01 Mar 2021

%\cite{Cipriani:2019aze}
\bibitem{Cipriani:2019aze}
M.~Cipriani [CMS],
%``Towards the $W$-boson mass measurement with the CMS experiment,''
Nuovo Cim. C \textbf{42}, no.4, 160 (2019).
%doi:10.1393/ncc/i2019-19160-4
%0 citations counted in INSPIRE as of 22 Feb 2021\end{thebibliography}

%\cite{Bertone:2017bme}
\bibitem{Bertone:2017bme}
V.~Bertone \textit{et al.} [NNPDF],
%``Illuminating the photon content of the proton within a global PDF analysis,''
SciPost Phys. \textbf{5}, no.1, 008 (2018).
%doi:10.21468/SciPostPhys.5.1.008
%[arXiv:1712.07053 [hep-ph]].
%89 citations counted in INSPIRE as of 23 Feb 2021

%\cite{Manohar:2017eqh}
\bibitem{Manohar:2017eqh}
A.~V.~Manohar, P.~Nason, G.~P.~Salam and G.~Zanderighi,
%``The Photon Content of the Proton,''
JHEP \textbf{12}, 046 (2017).
%doi:10.1007/JHEP12(2017)046
%[arXiv:1708.01256 [hep-ph]].
%117 citations counted in INSPIRE as of 23 Feb 2021

%\cite{Manohar:2016nzj}
\bibitem{Manohar:2016nzj}
A.~Manohar, P.~Nason, G.~P.~Salam and G.~Zanderighi,
%``How bright is the proton? A precise determination of the photon parton distribution function,''
Phys. Rev. Lett. \textbf{117}, no.24, 242002 (2016).
%doi:10.1103/PhysRevLett.117.242002
%[arXiv:1607.04266 [hep-ph]].
%194 citations counted in INSPIRE as of 23 Feb 2021


%\cite{Carrazza:2016htc}
\bibitem{Carrazza:2016htc}
S.~Carrazza, S.~Forte, Z.~Kassabov and J.~Rojo,
%``Specialized minimal PDFs for optimized LHC calculations,''
Eur. Phys. J. C \textbf{76}, no.4, 205 (2016).
%doi:10.1140/epjc/s10052-016-4042-8
%[arXiv:1602.00005 [hep-ph]].
%22 citations counted in INSPIRE as of 23 Feb 2021


\bibitem{Carrazza:2016wte}
S.~Carrazza and Z.~Kassabov,
%``Parton Distribution Functions at LHC and the SMPDF web-based application,''
PoS \textbf{PP@LHC2016}, 020 (2016).
%doi:10.22323/1.278.0020
%[arXiv:1606.09248 [hep-ph]].
%3 citations counted in INSPIRE as of 25 Feb 2021



%\cite{Giele:1998uh}
\bibitem{Giele:1998uh}
W.~T.~Giele and S.~Keller,
%``Determination of $W$ boson properties at hadron colliders,''
Phys. Rev. D \textbf{57}, 4433-4440 (1998).
%doi:10.1103/PhysRevD.57.4433
%[arXiv:hep-ph/9704419 [hep-ph]].
%49 citations counted in INSPIRE as of 23 Feb 2021

\end{thebibliography}
\end{document}